\documentclass[useAMS,usenatbib]{mn2e}

\usepackage{graphicx}
\usepackage{natbib}
\usepackage{longtable}

\usepackage{txfonts}
\newcommand{\asec}{$^{\prime\prime}$}

\title[Herschel Fornax Cluster]{The Herschel Fornax Cluster Survey I: The Bright Galaxy Sample}
\author[Davies et al.]
{J. I. Davies$^{1}$,
S. Bianchi$^{2}$,
M. Baes$^{3}$, 
A. Boselli$^{4}$,
L. Ciesla$^{4}$,
M. Clemens$^{5}$,
T. A. Davis$^{6}$, \newauthor
I. De Looze$^{3}$, 
S. di Serego Alighieri$^{2}$,
C. Fuller$^{1}$,
J. Fritz$^{3}$, 
L. K. Hunt$^{2}$,
P. Serra$^{7}$, \newauthor
M. W. L. Smith$^{1}$, 
J. Verstappen$^{3}$, 
C. Vlahakis$^{8}$, 
E. M. Xilouris$^{9}$,
D. Bomans$^{10}$, \newauthor
T. Hughes$^{11}$, 
D. Garcia-Appadoo$^{8}$
and S. Madden$^{12}$. \\ 
$^{1}$School of Physics and Astronomy, Cardiff University, The Parade, Cardiff, CF24
3AA, UK. \\
$^{2}$INAF-Osservatorio Astrofisico di Arcetri, Largo Enrico Fermi 5, 50125 Firenze, Italy. \\
$^{3}$Sterrenkundig Observatorium, Universiteit Gent, KrijgslAAn 281 S9, B-9000 Gent,
Belgium. \\
$^{4}$Laboratoire d'Astrophysique de Marseille, UMR 6110 CNRS, 38 rue F. Joliot-Curie,
F-13388 Marseille, France. \\
$^{5}$INAF-Osservatorio Astronomico di Padova, Vicolo dell'Osservatorio 5, 35122 Padova,
Italy. \\
$^{6}$European Southern Observatory, Karl-Schwarzschild Str. 2, 85748 Garching bei Muenchen, Germany.  \\
$^{7}$Netherlands Institute for Radio Astronomy (ASTRON),
Postbus 2, 7990 AA Dwingeloo, The Netherlands \\
$^{8}$Joint ALMA Observatory (JAO), Vitacura, Santiago, Chile. \\
$^{9}$Institute for Astronomy, Astrophysics, Space Applications \& Remote
Sensing, National Observatory of Athens, P. Penteli, 15236, Athens,
Greece. \\
$^{10}$Astronomical Institute, Ruhr-University Bochum, Universitaetsstr. 150, 44780 Bochum,
Germany. \\
$^{11}$Kavli Institute for Astronomy \& Astrophysics, Peking University, Beijing 100871,
P.R. China. \\
$^{12}$Laboratoire AIM, CEA/DSM- CNRS - Universit\'e Paris Diderot, Irfu/Service, Paris, France. \\
}

\begin{document}

\date{Original May 2012}


\maketitle


\begin{abstract}
We present Herschel Space Telescope observations of the nearby Fornax cluster at 100, 160, 250, 350 and 500$\mu$m with a spatial resolution of 7 - 36 arc sec (10"$\approx$1 kpc at $d_{Fornax}=$ 17.9 Mpc). We define a sample of eleven bright galaxies, selected at 500$\mu$m, that can be directly compared with our past work on the Virgo cluster. We check and compare our results with previous observations made by IRAS and Planck, finding good agreement. The far-infrared luminosity density is higher, by about a factor of three, in Fornax compared to Virgo, consistent with the higher number density of galaxies. The 100$\mu$m (42.5-122.5 $\mu$m) luminosity is two orders of magnitude larger in Fornax than in the local field as measured by IRAS. We calculate stellar ($L_{0.4-2.5}$) and far-infrared ($L_{100-500}$) luminosities for each galaxy and use this to estimate a mean optical depth of $\tau=0.4\pm0.1$ - the same value as we previously found for Virgo cluster galaxies. For ten of the eleven galaxies (NGC1399 excepted) we fit a modified blackbody curve ($\beta=2.0$) to our observed flux densities to derive dust masses and temperatures of $10^{6.54-8.35}$ M$_{\odot}$ and T=14.6-24.2K respectively, values comparable to those found for Virgo. The derived stars-to-gas(atomic) and gas(atomic)-to-dust ratios vary from 1.1-67.6 and 9.8-436.5 respectively, again broadly consistent with values for Virgo. Fornax is a mass overdensity in stars and dust of about 120 when compared to the local field (30 for Virgo). Fornax and Virgo are both a factor of 6 lower over densities in gas(atomic) than in stars and dust indicating loss of gas, but not dust and stars, in the cluster environment. We consider in more detail two of the sample galaxies. As the brightest source in either Fornax and Virgo, NGC1365 is also detected by Planck. The Planck data fit the PACS/SPIRE SED out to 1382$\mu$m with no evidence of other sources of emission ('spinning dust', free-free, synchrotron). 
At the opposite end of the scale NGC1399 is detected only at 500$\mu$m with the emission probably arising from the nuclear radio source rather than inter-stellar dust. 

\end{abstract}


\section{Introduction} 
The two nearest galaxy clusters to us are Virgo and Fornax. The NASA Extra-galactic Database (NED) gives the distance of M87 (the giant elliptical at the heart of Virgo) as 16.8 Mpc, while the similar galaxy at the heart of Fornax (NGC1399) lies at 17.9 Mpc. Both clusters are at high galactic latitude ($b_{Virgo}=74.5^{o}$, $b_{Fornax}=-53.6^{o}$) and so are ideal targets for extra-galactic studies at all wavelengths. Virgo and Fornax were observed at optical wavelengths (photographic) in the 1980s leading to the most complete catalogues of cluster members (Binggeli et al. 1985, Ferguson 1989). Fornax is somewhat less populous than Virgo with just 350 members listed in the Fornax Cluster Catalogue (FCC, Ferguson 1989) compared to about 2000 in the Virgo Cluster Catalogue (VCC, Binggeli et al. 1985).

\begin{figure*}
\includegraphics[scale=0.74]{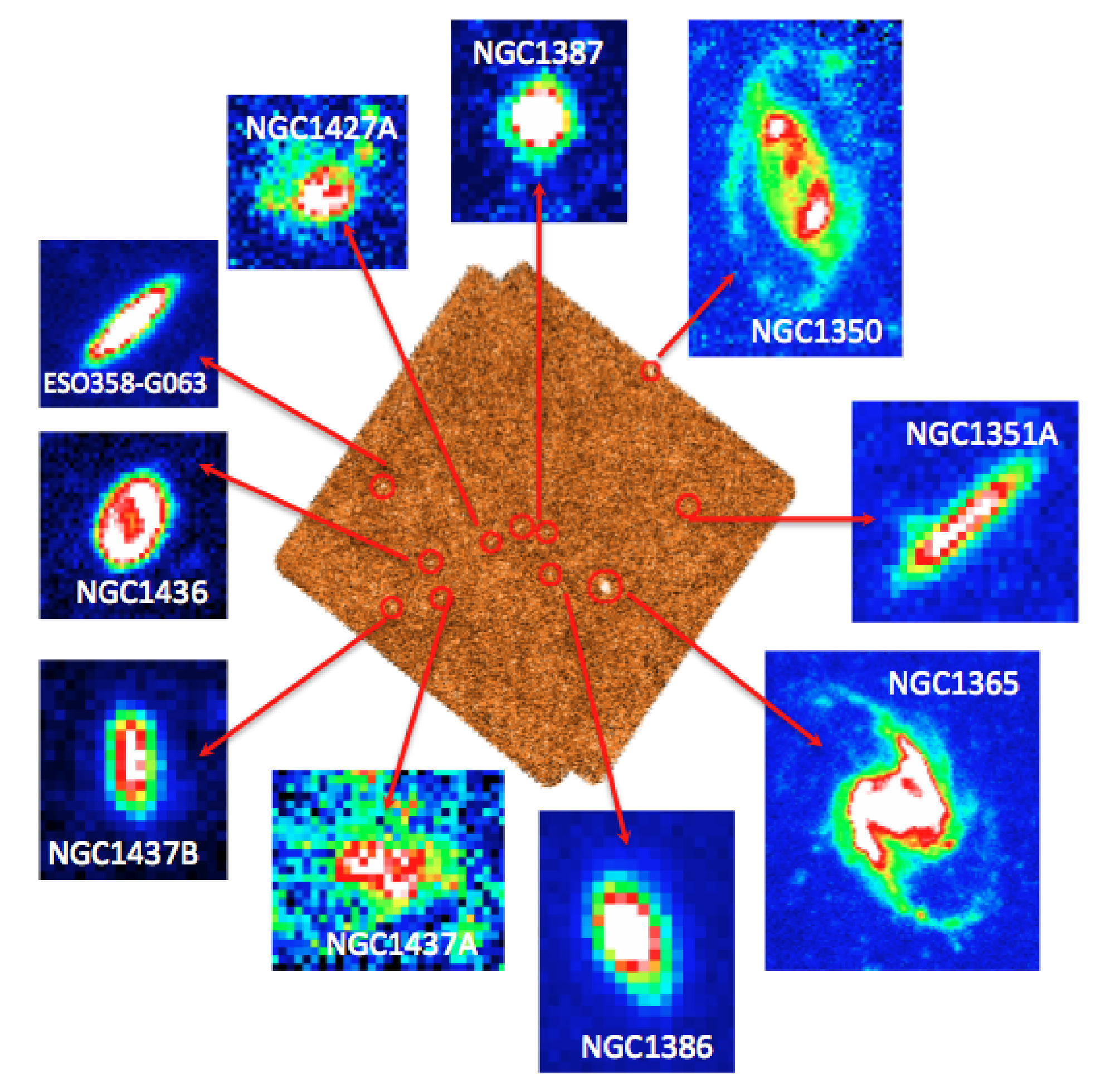}
\caption{The HeFoCS 500$\mu$m data with the positions and names of 10 of the 11 detected bright galaxies marked (the exception is NGC1399, which is shown in more detail in Fig. 5). Inset are 250$\mu$m images of 10 of the galaxies (NGC1399 is not detected at 250$\mu$m). } 
\end{figure*}

The Fornax cluster has a velocity dispersion of $\sim370$ km s$^{-1}$ and a core radius of 0.7 deg (250 kpc), which gives a mass of $\approx 7 \times 10^{13}$ M$_{\odot}$, about a factor of seven less than Virgo (Jordan et al. 2007 and references therein) and so Fornax might also be described as a large group. Like Virgo, Fornax does show evidence of sub-structure indicating its current and continuing assembly (Drinkwater et al. 2001) though it is more regular in shape and probably more dynamically evolved than Virgo. Although less massive than Virgo, the central galaxy number density is about three times higher in Fornax than in Virgo and with its lower velocity dispersion we might expect more influential interactions of Fornax galaxies with one another. To quantify this the relaxation time of Fornax, a measure of the time scale for gravitational interactions, is about one third that of Virgo ($T_{Relax}^{Virgo} \approx 4 \times 10^{10}$ years, Boselli and Gavazzi, 2006). However, this timescale is long compared to the crossing time which determines how long a mechanism like ram pressure stripping might be effective. The crossing time of Fornax and Virgo are about equal ($T_{Cross}^{Virgo} \approx 2 \times 10^{9}$ years, Boselli and Gavazzi, 2006), but the lower velocity dispersion of Fornax makes the inter-galactic medium less influential for the same inter-galactic medium density. We can quantify how 'effective' ($E$) ram pressure stripping might be using $E \propto t_{Cross} \delta v^{2} \rho_{g}$ (Gunn and Gott, 1972) where $\delta v$ is the cluster velocity dispersion and $\rho_{g}$ is the gas density. From the x-ray data (Scharf et al. 2005, Paolillo et al. 2002, Schindler et al. 1999) the central gas density and velocity dispersion of Virgo is about 4 and 2 times higher respectively than Fornax making Virgo potentially about 16 times more 'effective' at ram pressure stripping. 
So, because of these diverse cluster properties Virgo and Fornax galaxies may have quite different gas accretion/stripping histories. 

Our new observations of Fornax cluster galaxies described below make use of the unique imaging qualities and sensitivity of the Herschel Space Telescope at five wavelengths from 100 - 500$\mu$m. With a 3.5m mirror the spatial resolution of Herschel at these wavelengths ranges from 7 - 36\asec enabling spatially resolved observations of many of the bright galaxies at the distance of Fornax (10\asec$\equiv1$ kpc). 

The Herschel Fornax Cluster Survey (HeFoCS) is an ESA Herschel Space Observatory (Pilbratt et al., 2010) Open Time Project. The project was
awarded 31 hours of observing time in parallel mode using
PACS (Poglitsch et al., 2010) at  100 and 160$\mu$m and SPIRE (Griffin et al., 2010)
at 250, 350 and 500$\mu$m. Pre-Herschel comparable surveys describing the far infrared properties of nearby bright galaxies in both cluster and field have been described by Soifer et al. (1987), Doyon and Joseph (1989) (IRAS), Tuffs et al, (2002) (ISO), Draine et al. (2007) (Spitzer). Davies et al. (2012) have presented Herschel observations of the bright galaxies in the Virgo cluster.

In this paper we describe the first results from our Fornax cluster survey concentrating on the properties of the bright galaxies that are detected at high signal-to-noise. In many ways this paper is a continuation of the work we have previously done using the Herschel Virgo Cluster Survey (HeViCS) data, which is described in the following series of papers: {\bf Paper I} (Davies et al. 2010) that considered the properties of the bright galaxies in a single central 4$\times$4 sq deg HeViCS field, {\bf Paper II} (Cortese et al. 2010) that describes the truncation of cluster galaxy dust discs, {\bf Paper III} (Clemens et al. 2010) on the dust life-time in early-type galaxies, {\bf Paper IV} (Smith et al. 2010) considers the spiral galaxy dust surface density and temperature distribution, {\bf Paper V} (Grossi et al. 2010) on the properties of metal-poor star-forming dwarf galaxies, {\bf Paper VI} (Baes et al. 2010) looks at the lack of thermal emission from the elliptical galaxy M87, {\bf Paper VII}, (De Looze et al. 2010) discusses the far-infrared detection of dwarf elliptical galaxies, {\bf Paper VIII} (Davies et al. 2012) looks at the properties of the 78 brightest Virgo galaxies, {\bf Paper IX} (Magrini et al. 2011) compares galactic metalicity and dust-to-gas ratio gradients, {\bf Paper X} (Corbelli et al. 2012) considers the relationship between cold dust and molecular gas, {\bf Paper XI} (Pappalardo et al. 2012) assesses the environmental effects on molecular gas and dust in spiral disks and {\bf Paper XII} (Auld et al. 2012) discusses the detection of far-infrared emission from all catalogued VCC galaxies. A further paper (Boselli et al., 2010) discusses the spectral energy distributions of HeViCS galaxies together with  others observed as part of the Herschel Reference Survey (HRS). 

\section{Observations, data reduction, object selection and calibration checks}
We have obtained $\sim16$ sq deg of data from a single field centred on the central Fornax cluster galaxy NGC1399 (RA(J2000)=03:38:29.08, Dec(J2000)=-35:27:2.7), Fig. 1. We use parallel mode and a fast scan rate of 60\arcsec/sec (as used for the HeViCS data) over two$\times$two orthogonal cross-linked
scan directions (4 scans).

PACS data reduction was carried out with the standard pipeline for both the 100
and 160$\mu$m channels up to Level-1.
The four scans
were then combined and maps made using the {\it Scanamorphos} map-maker (Roussel, 2012). We chose to use {\it Scanamorphos}, instead of the phot-project mapper used in previous papers (Davies et al., 2012), as this avoids the necessity of high-pass filtering, which would remove the lower level  
extended emission. After combining orthogonal scans the full width half maximum (FWHM) beam sizes are approximately 7\arcsec and 13\arcsec with pixel sizes of 2\arcsec and 3\arcsec for the 100
and 160 $\mu$m channels respectively. 

The SPIRE data were also processed up to Level-1
with a custom driven pipeline script adapted from the official pipeline
({\sl
POF5\_pipeline.py}, dated 8 Jun 2010) as provided by the SPIRE Instrument
Control Centre (ICC)
\footnote{See 'The SPIRE Analogue Signal Chain and
Photometer Detector Data Processing Pipeline' Griffin et al. 2009 or
Dowell et al. 2010 for a more detailed description of the pipeline and a list
of the individual modules.}.
This {\it Jython} script was run in the Herschel Interactive Processing
Environment (HIPE - Ott, 2010). 
Our data reduction up to Level-1 is very
similar to
the Herschel Common Science System/Standard Product Generation v5 with a calibration based on Neptune data. 
Specific differences to the standard pipeline were that we used the {\it sigmaKappaDeglitcher} instead of the ICC-default 
{\it waveletDeglitcher}. Furthermore, we did not run the default {\it temperatureDriftCorrection} and the residual,
median baseline subtraction. Instead we use a custom method called BriGAdE (Smith et al. in preparation)
to remove the temperature drift and bring all bolometers to the same level (equivalent to baseline removal).
We have found this method improves the baseline subtraction significantly especially
in cases where there are
strong temperature variations during the observation.

Scans were then combined to make our final maps using the naive mapper provided in the standard pipeline.
The FWHM of the SPIRE beams are
18.2\arcsec, 24.5\arcsec, and 36.0\arcsec with pixel sizes of 6\arcsec,
8\arcsec, and 12\arcsec at 250, 350, and 500 $\mu$m, respectively. The beam areas used were 423, 751 and 1587 sq. arc sec respectively.
The final data products have a 1$\sigma$ noise, determined from the whole of each image, of $\sim$0.6, 0.7, 1.0, 1.0 and 1.0 mJy pixel$^{-1}$ or 6.4, 3.3, 1.2, 0.7 and 0.3 MJ sr$^{-1}$ at 100, 160, 250, 350 and 500 $\mu$m respectively. 

As in Davies et al. (2012) for the Virgo cluster we obtain a bright far-infrared selected sample by carrying out our initial object selection at 500 $\mu$m. This is because it is the least explored part of the spectrum, it has the lowest resolution, and most galaxies will produce their lowest flux in this band, which should guarantee a detection in all five bands. 
To produce an objectively selected sample we used the automatic image detection algorithm SExtractor (Bertin and Arnouts, 1996). To minimise background contamination by faint sources each object was required to have more than 41 connected pixels at 1.5$\sigma_{500}$ or above \footnote{We used 30 connected pixels for the Virgo (HeViCS) data in Davies et al. (2012), but since then we have changed our pixel scale from 14 to 12 arc sec at 500$\mu$m and so the two are equivalent.}. The result is a 500 $\mu$m flux density limit of $\sim$0.1 Jy for sources with a diameter larger than 1.4\arcmin. Each object was then checked for correspondence with a known Fornax Cluster galaxy (FCC). The final sample consists of 11 Fornax Cluster objects, which is just 4\% of the 256 FCC galaxies in our field. This compares with 78 galaxies detected in just the same way in the Virgo cluster, which is 12\% of the galaxies in this area listed in the VCC (Davies et al. 2012). A more straight forward comparison is that 14 galaxies were detected in the central 16 sq deg Virgo field centred on M87, compared to 11 over the same area centred on NGC1399. Discarded objects are in the background of Fornax. 

In the standard way the data were initially smoothed and re-gridded to the 500 $\mu$m resolution 
and pixel scale. Elliptical apertures were chosen by eye using the 500$\mu$m data, with the sky defined by a concentric annulus. These same annuli were then used on the smoothed and re-gridded data at other wavelengths. As a test we compared this method with carrying out the aperture photometry on the original unsmoothed 100$\mu$m data, our criteria being correspondence with the flux density values at 100$\mu$m from IRAS. These tests showed that we obtained a much better result by using the original resolution data, rather than the smoothed and re-gridded data.
So, independent measurements were carried out on the maps at the original resolution,
defining by eye for each galaxy an elliptical aperture and a nearby sky aperture that avoided pixels 
contaminated by bright background sources. The flux densities quoted in table 1 come from aperture photometry using this unsmoothed and original resolution data.

\begin{figure}
\centering
\includegraphics[scale=0.5]{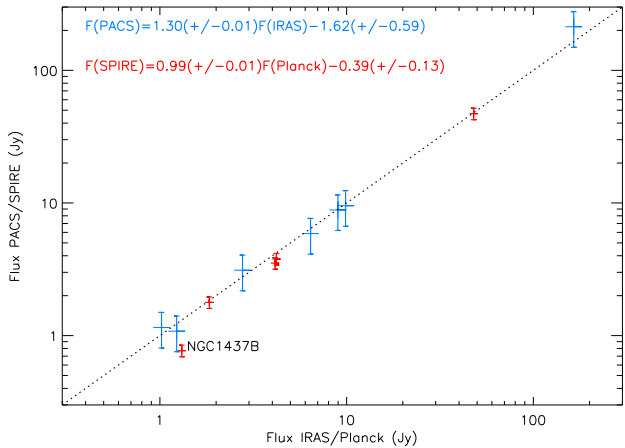}
\caption{A comparison of IRAS (blue) flux densities with the PACS 100$\mu$m and the Planck (red) flux densities with the SPIRE 350$\mu$m. Linear least squares fit parameters are given top left. The black dashed line is the one to one relationship.} 
\end{figure}

\begin{table*}
\begin{center}
\begin{tabular}{lccccccccc} \hline
(1) & (2) & (3) & (4) & (5) & & & (6) & & \\ 
Name    &  RA        &   Dec    & $v$           & $d_{Mpc}$     & $F_{500}$ & $F_{350}$  & $F_{250}$  & $F_{160}$  & $F_{100}$ \\ 
        & (J2000)    &  (J2000) & (km s$^{-1}$) & (Mpc) & (Jy) & (Jy) & (Jy) & (Jy) & (Jy) \\ \hline
NGC1351A    &  03:28:48.8 &	-35:10:42.2  & 1353 & 20.9$\pm1.9$ &  0.41  &  0.91  &  1.71  &  2.32  &  1.15 \\
NGC1350     &  03:31:05.6 &	-33:38:27.5  & 1905 & 20.9$\pm3.6$ &  1.51  &  3.78  &  7.15  &   -    &   -   \\
NGC1365     &  03:33:35.9 &	-36:08:25.9  & 1636 & 17.9$\pm2.7$ &  17.66 &  47.14 & 106.98 & 205.32 & 213.06 \\
NGC1386     &  03:36:46.4 &	-35:59:57.1  & 868  & 16.2$\pm0.8$ &  0.59  &  1.57  &  3.93  &  8.45  &  8.86  \\
NGC1387     &  03:36:56.8 &	-35:30:27.4  & 1302 & 17.2$\pm3.1$ &  0.32  &  1.09  &  2.80  &  6.10  &  5.88  \\
NGC1399     &  03:38:27.0 &	-35:26:46.1  & 1429 & 17.9$\pm2.3$ &  0.26  &   -    &   -    &   -    &   -  \\
NGC1427A    &  03:40:08.2 &	-35:37:32.5  & 2028 & 16.4$\pm0.7$ &  0.22  &  0.43  &  0.69  &  1.07  &   -  \\
NGC1437A    &  03:43:02.7 &	-36:16:17.1  & 886  & 16.9 &  0.23  &  0.31  &  0.44  &  0.62  &   -  \\
NGC1436     &  03:43:37.0 &	-35:51:12.7  & 1387 & 19.1$\pm1.3$ &  0.66  &  1.78  &  4.07  &  5.45  &  3.11 \\
NGC1437B    &  03:45:54.1 &	-36:21:34.6  & 1497 & 10.3 &  0.39  &  0.77  &  1.39  &  1.78  &  1.08 \\
ESO358-G063 &  03:46:18.9 &	-34:56:40.9  & 1929 & 18.9$\pm4.2$ &  1.39  &  3.52  &  7.64  & 12.66  &  9.53 \\ \hline
\end{tabular}
\caption{The Herschel Fornax Cluster Survey Bright Galaxy Sample - (1) name, (2) (3) position, (4) velocity, (5) distance, and (6) far-infrared flux density. A dash indicates a non-detection by Herschel or in the case of NGC1350 it is not observed by PACS. We estimate total uncertainties in our flux density values of 25, 15, 10, 10 and 15\% at 100, 160, 250, 350 and 500$\mu$m respectively.}
\end{center}
\end{table*}

Flux density errors are estimated as in Davies et al. (2012). A large part of the 
uncertainty is due to the background subtraction: using different sized
apertures for the same object introduces an error of
17, 10, 8, 8 and 10\% at 100, 160, 250, 350 and 500$\mu$m respectively.
A second source of error, which is large for PACS data, is
due to data recording (the repeatability of independent observations of the same source) and image making (aligning, scaling and filtering). We found it to be 20, 10, 2, 3 
and 5\%. Finally, we include the calibration error, which is claimed
to be below 5\% for PACS (Mueller et al. 2011) and 7\% for SPIRE
(Swinyard et al., 2010). Taking the above three uncertainties to
be independent we estimate total uncertainties in our flux values of 27,
15, 11, 11 and 13\% at 100, 160, 250, 350 and 500$\mu$m respectively.

\begin{table*}
\begin{center}
\begin{tabular}{l|cccc}
Band    &  Mean luminosity &  Luminosity density & Mean luminosity &  Luminosity density \\
($\mu$m) & (Virgo)         &      (Virgo)        &   (Fornax)      &      (Fornax)      \\
         & $\times 10^{22}$ (W Hz$^{-1}$ sr$^{-1}$) & $\times 10^{-45}$ (W m$^{-3}$ Hz$^{-1}$) & $\times 10^{22}$ (W Hz$^{-1}$ sr$^{-1}$) & $\times 10^{-45}$ (W m$^{-3}$ Hz$^{-1}$) \\ \hline
100 & 3.4  & 6.5(19.7) & 9.4 & 24.5(39.9) \\
160 & 4.4  & 8.3(25.9) & 8.3 & 25.2(40.3) \\
250 & 2.4  & 4.6(14.8) & 4.3 & 14.2(23.1) \\
350 & 1.0  & 1.9(6.3)  & 1.9 & 6.1(9.9)  \\
500 & 0.4  & 0.7(2.3)  & 0.6 & 2.2(3.6) \\
\end{tabular}
\caption{The mean luminosity and luminosity density in each band for the Virgo and Fornax bright galaxy samples. The Virgo luminosity density is that for the whole sample which lies over a distance range of 17-32 Mpc. Values in brackets are for the sample of galaxies with distances of 17-23 Mpc (see Davies et al. 2012). The Fornax values are for the whole sample with values in brackets with NGC1437B, which has a somewhat discrepant distance (10.3 Mpc), removed. Errors on these values are difficult to assess because of the unknown distance errors and our definition of the cluster volume. Errors on the measured flux densities are given in the text.}
\end{center}
\end{table*}

In table 1 we list the Fornax bright galaxies with their names, co-ordinates, velocities, distances and flux densities in each band. Co-ordinates are obtained from centroiding the 500$\mu$m data. The flux densities in Table 1 are as measured and do not include color corrections (see Section 5 of Davies et al. 2012). Velocities and distances have been taken from the NED. The value we have used for the distance is the mean redshift independent distance given in NED, where available we also include the standard deviation of the measurements. The number of measurements contributing to the mean range from 39 for NGC1365 to just 1 for NGC1437A and NGC1437B. Distances are derived in a number of different ways: primarily surface brightness fluctuations and planetary nebulae luminosity function for early type galaxies and Tully-Fisher for later types. Where required in the calibration of the method a Hubble constant of 75 km s$^{-1}$ Mpc$^{-1}$ has been used. Only NGC1365 has a cepheid distance obtained from the 'Final Results from the Hubble Space Telescope Key Project to Measure the Hubble Constant' (Freedman et al. 2001) and this agrees with the NED mean distance of 17.9 Mpc.

For both the PACS and SPIRE data we can make some comparisons of our results with those obtained by others. IRAS 100$\mu$m fluxes are available from NED for 10 of the 11 galaxies (the exception being NGC1437A which   
has not been detected at 100$\mu$m by HeFoCS nor IRAS). NGC1350 is not on the 100$\mu$m map and we do not detect NGC1399 or NGC1427A (NGC1399 is discussed in more detail in section 3). This leaves 7 galaxies that we can compare with IRAS as shown in Fig 2. Individual measurements are consistent between PACS and IRAS within the errors, but the slope of the best fit line is a steep 1.3. Removing the brightest galaxy NGC1365 ($F^{PACS}_{100}=213.06$, $F^{IRAS}_{100}=164.52$) from the sample leads to a slope of 0.95 - it is not clear from just this one galaxy whether this is a generic problem with bright sources or just specific to NGC1365, though Aniano et al. (2012) highlight a similar problem when relating PACS and Spitzer data. The IRAS/PACS colour corrections are negligible. NED gives ISO 150$\mu$m and 170$\mu$m flux densities of 194 and 167 Jy respectively for NGC1365. The mean of these is 181 Jy which compares with our value of $\sim205$ Jy at 160$\mu$m, again consistent within our calculated errors.  The Planck consortium have released a point source catalogue of bright sources which, at 350$\mu$m, contains 5 of the galaxies in our list. In Fig 2. we also compare these 350$\mu$m flux densities (we use the Planck GAUFLUX value for extended sources) with our SPIRE data - there is a reasonably close agreement between the SPIRE and Planck values, with only NGC1437B being slightly discrepant (see Fig. 2).  The SPIRE/Planck colour correction is small ($F_{350}^{SPIRE}/F_{350}^{Planck}$=0.99) and has been ignored, Ciesla et al. (2012). For NGC1365 we also have a Planck detection at 550$\mu$m of 13.62 Jy. For a typical dust temperature of 20K a correction of $\sim1.2$ is necessary to transform this to what is expected at 500$\mu$m (Baes et al., 2012). Carrying out this correction leads to a value of 16.34 Jy compared to 17.66 Jy measured from our SPIRE data, this value is again consistent within our quoted errors - the SED of NGC1365 is discussed further in section 3.

\section{The derived parameters of the sample galaxies}
\subsection{Luminosity and luminosity density}

Given the apparent shape (peaked) of the Virgo cluster luminosity distributions we decided to characterise each distribution solely by its mean values (Davies et al. 2012).  With so few galaxies in this sample this is also about all we can justifiably do for Fornax.  Mean values are given in table 2, where they are also compared to Virgo. It is clear that the majority of the far-infrared energy of the cluster is being produced at the shortest of these wavelengths. The mean luminosity of galaxies in Fornax is higher than that of Virgo because of the total dominance in the far-infrared of NGC1365. For example at 250$\mu$m 76\% of the luminosity of the cluster is produced by just this one galaxy.

For Virgo the luminosity distributions turned over at both the faint and luminous ends and so we could make an estimate of the luminosity density in each band. Based on the same assumption (we have too few Fornax galaxies to conclusively demonstrate this) we can in the same way calculate a luminosity density for Fornax. The volume sampled over 16 sq deg of sky is $\sim$12.9 Mpc$^{3}$ for galaxies between the extremes of distance of 10.3 and 20.9 Mpc. This seems a little large for a cluster, in which case it is the distance of NGC1437B that is discrepant. Removing NGC1437B from the sample the volume is $\sim$7.9 Mpc$^{3}$ (galaxies between 16.2 and 20.9 Mpc). Densities calculated with NGC1437B removed from the sample will be given in brackets after the value for the whole sample. As explained in Davies et al. (2012) Virgo cluster members in the HeViCS sample extend from 17 to 32 Mpc, which is also a large distance compared to what is normally assumed for the size of a cluster. So, all densities quoted for Virgo will be given using both the whole sample and using only those with distances between 23 and 17 Mpc given in brackets. 

It is not easy to define the size of a given cluster so that realistic comparisons can be made between clusters of different sizes. Ferguson (1989a) gives the core radii of Virgo and Fornax as $\approx0.6$ and $\approx0.3$ Mpc respectively. This is a ratio of a factor of four in area, just the same as the ratio of field sizes between the HeViCS and HeFoCS surveys and so we will compare cluster properties using galaxies detected in both cases over the full survey area.

Luminosity densities calculated in the above way are given in Table 2. Far-infrared luminosity densities are higher by about a factor of 3 in Fornax compared to Virgo. This is consistent with the factor of 2.5 times larger central number density of galaxies in Fornax compared to Virgo (Ferguson, 1989). Comparisons of luminosity densities with other surveys are discussed by Davies et al. (2012), but of particular interest is the Saunders et al. (1990) value for the mean local luminosity density. In the range 42.5-122.5$\mu$m they derive a value of $4.0 \times 10^{7}$ L$_{\odot}$ Mpc$^{-3}$. Using this band pass and our 100$\mu$m luminosity density we can make an approximate comparison with the Saunders et al. (1990) value\footnote{Note the Saunders et al. value is derived from both IRAS 60 and 100$\mu$m flux densities. The 60$\mu$m flux density in particular may be influenced by a warmer dust component associated with star formation that is not included in the fits to our data}. We calculate a value of $5.8(8.6) \times 10^{9}$ L$_{\odot}$ Mpc$^{-3}$ for Fornax compared with $1.1(3.3) \times 10^{9}$ L$_{\odot}$ Mpc$^{-3}$ for Virgo. The Fornax value is 145(238) times larger than the local mean value measured by IRAS.

The far-infrared luminosity of each galaxy can be calculated using non-overlapping bands corresponding to each Herschel wavelength (not necessarily centred on the nominal wavelength because of the uneven band spacing). As in Davies et al. (2012) we define the Herschel far-infrared luminosity as: \\
\begin{center}
$L_{100-500}=3.1\times10^{4}d_{Mpc}^{2}[(f_{100}\Delta f_{100})+(f_{160}\Delta f_{160})+(f_{250}\Delta f_{250})$ \\
$+(f_{350}\Delta f_{350})+(f_{500}\Delta f_{500})]$ \hspace{0.5cm} $L_{\odot}$
\end{center}
where $f_{100}$, $f_{160}$, $f_{250}$, $f_{350}$, $f_{500}$, $\Delta f_{100}=18.0$, $\Delta f_{160}=8.9$, $\Delta f_{250}=4.6$, $\Delta f_{350}=3.1$ and $\Delta f_{500}=1.8$ are the flux density (Jy) and bandwidth ($10^{11}$ Hz) in each band respectively and we have taken the solar luminosity to be $3.9\times10^{26}$ W. Calculating the luminosity in this way means that it does not depend on any particular fit to the SED i.e. it is valid for galaxies well fitted by a single modified blackbody with any value of emissivity power index ($\beta$), those that require two components or more and those that do not have a thermal spectrum. 
Far-infrared luminosities ($L_{100-500}$) for each galaxy are listed in Table 3. Comparing the average far-infrared SED of the galaxies in the sample, we find the highest flux density occurs almost equally in the 100 and 160$\mu$m bands, with the flux density about one third of the total in each case. 

\begin{table*}
\begin{center}
\begin{tabular}{lcccccccccc} \hline
(1) & (2) & (3) & (4) & (5) & (6) & (7) & (8) & (9) & (10) & (11)\\ 
Name    & M$_{B}$ & Log(M$_{Stars}$)  & Log(M$_{HI}$)  & Log(M$_{Dust}$)  & T$_{d}$ 
&  Log(L$_{0.4-2.5}$) & Log(L$_{100-500}$) & $<\tau>$ & M$_{Stars}$/M$_{HI}$ & M$_{HI}$/M$_{Dust}$ \\
       &          &    (M$_{\odot}$)  & (M$_{\odot}$)  & (M$_{\odot}$)    & (K)
&    (L$_{\odot}$) & (L$_{\odot}$)  & & &  \\ \hline           
NGC1351A    &  -17.39 &  9.41  & 8.80  & 7.04$\pm0.07$ & 17.1$\pm1.1$ &  8.93  & 8.85  & 0.61  & 4.1  & 57.5  \\
NGC1350     &  -20.44 &  10.71 & 9.15  & 7.68$\pm0.15$ & 16.8$\pm2.0$ &  10.26 &  -    &  -    & 36.3 & 29.5  \\
NGC1365     &  -21.03 &  10.87 & 10.08 & 8.35$\pm0.07$ & 22.0$\pm2.6$ &  10.50 & 10.80 &  1.10 & 6.2  & 53.7  \\
NGC1386     &  -18.96 &  10.24 & -     & 6.73$\pm0.07$ & 23.6$\pm2.8$ &  9.76  & 9.33  &  0.32 & -    &  - \\
NGC1387     &  -19.50 &  10.55 & -     & 6.58$\pm0.07$ & 24.2$\pm2.3$ &  10.00 & 9.20  &  0.15 & -    &  - \\
NGC1399     &  -20.71 &  11.05 & -     & -    & -      &  10.55 &  -    &   -   & -    & -  \\
NGC1427A    &  -17.63 &  9.23  & 9.18  & 6.54$\pm0.10$ & 16.4$\pm1.8$ &  9.23  & 8.08  &  0.09 & 1.1  & 436.5  \\
NGC1437A    &  -17.23 &  9.07  & 8.70  & 6.56$\pm0.13$ & 14.8$\pm$ &  9.07  & 7.89  &  0.06 & 2.3  & 138.0 \\
NGC1436     &  -19.00 &  10.00 & 8.17  & 7.18$\pm0.07$ & 18.5$\pm1.2$ &  9.54  & 9.15  &  0.35 & 67.6 & 9.8  \\
NGC1437B    &  -16.08 &  8.85  & 8.08  & 6.56$\pm0.08$ & 14.6$\pm1.0$ &  8.29  & 8.15  &  0.55 & 5.9  & 33.1  \\
ESO358-G063 &  -18.78 &  9.91  & 9.26  & 7.38$\pm0.07$ & 19.8$\pm1.3$ &  9.45  & 9.49  &  0.74 & 4.5  & 75.9  \\ \hline
\end{tabular}
\caption{The Fornax Cluster Survey Bright Galaxy Sample - (1) name, (2) absolute B magnitude, (3) stellar mass, (4) HI mass, (5) dust mass, (6) dust temperature, (7) stellar luminosity from 0.4 to 2.5$\mu$m, (8) far-infrared luminosity from 100 to 500$\mu$m, (9) mean optical depth, (10) stellar to atomic gas mass ratio and (11) atomic gas to dust mass ratio. Optical and near-infrared magnitudes where required are taken from the NED database. NGC1427A and NGC1437A were not detected by 2MASS and so their stellar mass has been calculated using just their B band magnitude and an absolute magnitude of the Sun of $M^{B}_{\odot}=5.47$ (for the same reason they do not have an integrated optical luminosity). NGC1399 only has a measurement at 500$\mu$m and so we are not able to calculated a dust mass. The early type galaxies NGC1386, NGC1387 and NGC1399 have not been detected at 21cm. }
\end{center}
\end{table*}

Simply multiplying the luminosity densities given in Table 2 by the band widths, as above, we obtain a Fornax 100-500$\mu$m far-infrared luminosity density of $5.8(9.5)\times10^{9}$ L$_{\odot}$ Mpc$^{-3}$ using a solar luminosity of $3.9\times10^{26}$ W. This is a little over 3 times higher than the Virgo value of $1.6(7.0)\times10^{9}$ L$_{\odot}$ Mpc$^{-3}$ (Davies et al. 2012).  

In Davies et al. (2012) we define the apparent stellar luminosity from 0.4 to 2.5$\mu$m as: \\
\begin{center}
$L_{0.4-2.5}=3.1\times10^{7}d_{Mpc}^{2}[(f_{g}\Delta f_{g})+(f_{r}\Delta f_{r})+(f_{i}\Delta f_{i})$ \\
$+(f_{J}\Delta f_{J})+(f_{H}\Delta f_{H}+(f_{K}\Delta f_{K}))]$ \hspace{0.5cm} $L_{\odot}$
\end{center}
where $f_{g}$, $f_{r}$, $f_{i}$, $f_{J}$, $f_{H}$, $f_{K}$, $\Delta f_{g}=1.9$, $\Delta f_{r}=1.1$, $\Delta f_{i}=1.6$, $\Delta f_{J}=0.9$, $\Delta f_{H}=0.3$ and $\Delta f_{K}=0.5$ are the flux (Jy) and bandwidth ($10^{14}$ Hz) in each band respectively. The $g$, $r$ and $i$ band data ($\lambda_{g}=0.48\mu m$, $\lambda_{r}=0.62\mu m$, $\lambda_{i}=0.76\mu m$) were taken from the Sloan Digital Sky Survey (SDSS). For the Fornax galaxies we do not have SDSS data or any other uniform optical data and so we have had to rely on the near infrared data only. In Davies et al. (2012) we showed that about 73\% of the observed stellar radiation is emitted in these NIR bands and so we have multiplied the calculated stellar luminosity by a factor of 1.36 to account for this. For NGC1427A and NGC1437A we do not have NIR data and so we have just used the blue band magnitude and an absolute blue band magnitude of the Sun of 5.45. In the near infrared the values used are from the two Micron All Sky Survey (2MASS) which lists J, H and K band (total) magnitudes ($\lambda_{J}=1.25\mu m$, $\lambda_{H}=1.65\mu m$, $\lambda_{K}=2.17\mu m$). The 2MASS website gives the following zero points for the conversion of magnitudes to Jy: $K_{J}=8.01$, $K_{H}=7.53$, $K_{K}=7.06$.

Summing the contribution from all the galaxies leads to an optical luminosity density of $\rho_{0.4-2.5} = 8.6(14.1)\times10^{9}$ L$_{\odot}$ Mpc$^{-3}$ just 1.5 times larger than the far-infrared value given above. The optical luminosity (uncorrected for extinction) is found to be about 3 times that of the far-infrared in Virgo (Davies et al. 2012). 

The optical and far-infrared luminosities can be used to make a crude estimate of the 'typical' optical depth ($<\tau>$) experienced by a photon as it leaves a galaxy, based on a simple screen of dust model:
\begin{center}
$<\tau>=\ln{\left(1.0+\frac{L_{100-500}}{L_{0.4-2.5}}\right)}$ \\
\end{center}
Values of $<\tau>$ are listed in Table 3. 
The mean value for galaxies in this sample is $<\tau>_{mean}=0.4\pm0.1$ so on average the optical energy is emerging from regions of intermediate optical depth (neither totally optically thin or thick) - the mean value found for Virgo was just the same. Individual values of $<\tau>$ also have a similar range to Virgo of 0.06 to 1.10. Interestingly NGC1365, the brightest far-infrared source in both Virgo and Fornax has a far infrared luminosity almost the same as its optical luminosity  and the typical photon originates from an optically thick ($\tau=1.11$) region. As pointed out in Davies et al. (2012), the two 'optically thick' galaxies with $<\tau>$ greater than unity in the Virgo sample are both relatively face-on late type spirals - NGC4234 (Sc) and NGC4299 (Scd); this is also now true for Fornax (inclinations to the plane of the sky are 56, 40 and 20$^{o}$ for NGC1365, NGC4234 and NGC4299 respectively). As galactic dust is typically confined to a relatively thin disc, the value of $\tau$ should be dependent on the inclination of each galaxy to the line of sight. It is therefore surprising that the highest values of $\tau$ are associated with relatively face-on galaxies. The above value of $<\tau>_{mean}$ is very close to the value given in Saunders et al. (1990) derived in a similar way using IRAS data ($<\tau>_{mean}^{IRAS} = 0.3\pm0.1$).
The value we find implies that on average $\sim33$ \% of the stellar radiation of a galaxy is absorbed by dust. The value is in
agreement with what was found by Popescu and Tuffs (2002) using a sample of late-type Virgo galaxies observed with ISO.

\subsection{Dust mass and temperature}
In order to derive dust mass and temperature, we have fitted the
SED of each galaxy, as defined by the flux densities given in Table 1,
with a single temperature modified blackbody. The method is identical to that used for the Virgo galaxies and is fully described in Davies et al. (2012). In summary, we use a power 
law dust emissivity $\kappa_\lambda=\kappa_0 (\lambda_0/\lambda)^\beta$,
with spectral index $\beta=2$ and emissivity $\kappa_0$ = 0.192 $m^2$ kg$^{-1}$
at $\lambda_0$ = 350$\mu$m. The monocromatic flux densities measured from
the images, as reported in Table 1, come from the pipeline
calibration. They have been derived from the passband-weighted flux 
density (measured by the instruments), applying a color correction
for a flat energy spectrum ($F_\nu \propto \nu^{-1} $). When doing the fit to the data points we first remove this correction and then apply our own color correction to the
pipeline monochromatic fluxes densities.
The color corrections are given in Table 3 of Davies et al. (2012).
The model fit to the data was obtained with a standard $\chi^2$ 
minimization technique. 

\begin{figure}
\includegraphics[scale=0.575]{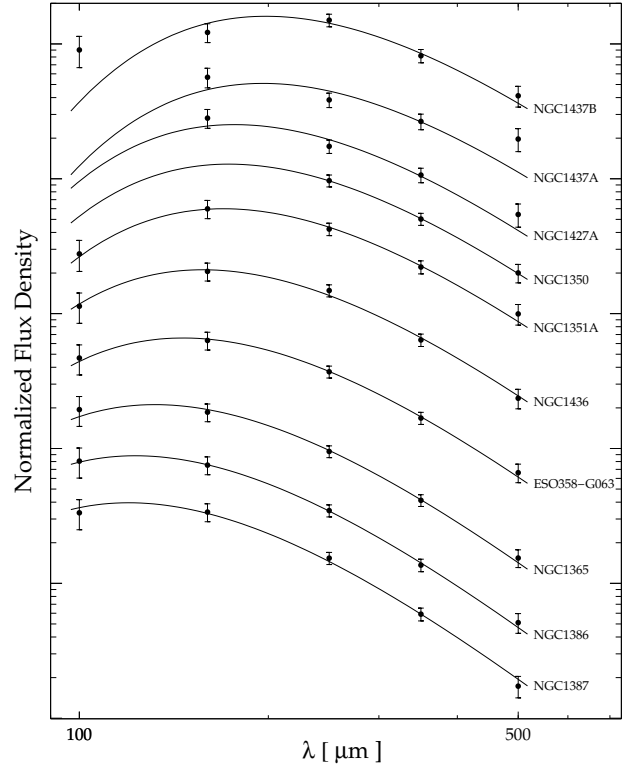}
\caption{Herschel spectral energy distributions of the Fornax bright galaxy sample. The SEDs are ordered with the coldest galaxy at the top.} 
\end{figure}

Plotting the data given in Table 1 shows that the majority of galaxies are simply and well fitted by this single temperature modified blackbody curve with fixed emissivity index; the fits are shown in Fig. 3 (ordered by increasing temperature). This good correspondence of the data with a single modified blackbody curve is essentially the same result we obtained for Virgo (Davies et al. 2012, see also Boselli et al. 2012 and Dale et al. 2012). The most notable discrepant point is that at 100$\mu$m for NGC1437B. As with many Fornax galaxies when compared to Virgo, there is no literature discussion of NGC1437B that we can find. It is described as an I0 galaxy in NED and as an edge-on Sd in Thomas et al. (2008). There is a GALEX image that shows that there is star formation in bright 'clumps' and it clearly looks as though it has been disturbed in some way by the local environment. Presumably the excess 100$\mu$m emission is due to warm dust heated by this on-going star formation. NGC1427A and NGC1437A both have what might be excess emission at 500$\mu$m, but it is difficult to be sure because there are only 4 data points and the modified blackbody fit is not so good. Note that NGC1399 is omitted from Fig. 3 because we only have one flux density measurement at 500$\mu$m.

Derived dust masses and temperatures are in the range $10^{6.54-8.35}$ M$_{\odot}$ and 14.6-24.2K respectively ($10^{6.22-8.17}$ M$_{\odot}$ and 12.8-27.2K for Virgo).  Prior to the availability of observations at wavelengths longer than about 100$\mu$m, calculated galaxy dust masses and temperatures from surveys similar to HeViCS and HeFoCS were typically $10^{6.6}$ M$_{\odot}$ and 30-50K respectively (taken from Soifer et al., 1987 where they have 31 galaxies in common with HeViCS, see also Devereux and Young, 1990). This confirms the need for longer wavelength observations ($\lambda >
160$$\mu$m) to sample the Rayleigh-Jeans tail of the SED and detect the colder
dust component in galaxies.

\subsection{Stellar mass}
We have calculated stellar masses ($M_{Star}$) for each galaxy (Table 3) using the prescription given in Bell et al. (2003) i.e.
\begin{center}
$\log{M_{star}}=-0.359+0.21(B-V)+\log{\frac{L_{H}}{L_{\odot}}}$
\end{center}
$L_{H}$ has been calculated using a H band absolute magnitude for the Sun of M$^{\odot}_{H}$=3.32. $M_{H}$ and (B-V) values have been taken from NED. Because of the lack of data, for NGC1427A and NGC1437A the stellar masses have been calculated solely from their B band magnitudes using a B band magnitude for the Sun of M$^{\odot}_{B}$=5.47.

\begin{table*}
\begin{center}
\begin{tabular}{l|ccccc}
      &  Virgo                     &  Fornax                    & Field & Virgo  & Fornax \\
      &  (M$_{\odot}$ Mpc$^{-3}$)  & (M$_{\odot}$ Mpc$^{-3}$)   & (M$_{\odot}$ Mpc$^{-3}$) & overdensity & overdensity   \\ \hline
Stars & $7.8(29.7) \times 10^{9}$      &  $24.0(39.2) \times 10^{9}$     & $0.2\pm0.1 \times 10^{9}$ & 34(129)  & 120(195)  \\
Gas   & $0.5(1.4) \times 10^{9}$   &  $1.4(2.3) \times 10^{9}$  & $0.08\pm0.01 \times 10^{9}$ & 6(18) & 18(29)  \\
Dust  & $8.6(27.8) \times 10^{6}$   & $26.5(43.3) \times 10^{6}$  & $0.22\pm0.04 \times 10^{6}$  & 39(126) & 120(196) \\
\end{tabular}
\caption{The mass densities in stars, gas and dust in the Fornax and Virgo clusters along with that for the local field. represent when compared to the local field. The Virgo values are for the whole sample, which lies over a distance range of 17-32 Mpc. Values in brackets are for the sample of galaxies with distances of 17-23 Mpc (see Davies et al. 2012). The Fornax values are for the whole sample with values in brackets with NGC1437B, which has a somewhat discrepant distance (10.3 Mpc), removed. References for the field values are given in the text. The final two columns give the over densities of Virgo and Fornax compared to the field value.}
\end{center}
\end{table*} 

\subsection{Gas mass}
Where available (8 out of 11 galaxies), we have taken the 21cm spectral line flux integral ($S_{Tot}$, Jy km s$^{-1}$) for each galaxy from NED and used the distances given in Table 1 to obtain atomic hydrogen gas masses ($M_{HI}$, M$_{\odot}$). The masses have been obtained using the standard formula:
\begin{center}
$M_{HI}=2.4 \times 10^{5} d^{2}_{Mpc} S_{Tot}$
\end{center}
Atomic hydrogen gas masses are given in Table 3 along with dust and stellar mass.

Derived from CO observations there are 4 detections and 3 upper limits on the molecular hydrogen masses of these galaxies (Horellou et al. 1995). NC1365, NGC1386, NGC1436 and ESO358-G063 have Log($M_{H_{2}}$) equal to 10.3, 8.3, 7.7 and 8.6 respectively. There is almost a factor of 2 more molecular hydrogen in NGC1365 than atomic and NGC1386 is surprisingly detected in CO, but not HI. NGC1351A, NGC1350 and NGC1387 all have upper limits on their molecular gas masses that are in each case more than a factor of 10 less than their atomic hydrogen masses. We have not incorporated this molecular gas data into Table 3 or discussed it further so that this paper is directly comparable with the equivalent paper on Virgo, where we have not included molecular gas masses (Davies et al. 2012). In the above we have taken $X_{CO}=3 \times 10^{20}$ cm$^{2}$ (K km/s)$^{-1}$.

\subsection{Mass densities}
By simply dividing the total mass in each component by the volume sampled ($\sim$12.9(7.2) Mpc$^{3}$) we can calculate the cluster mass densities. Provided the luminosity distributions in each far-infrared band are peaked the total cluster dust mass should be well constrained by our sample galaxies unless there is a significant dust component in the inter-galactic medium. From the dust in our sample galaxies we derive the densities given in Table 4. Values for Virgo have been taken from Davies et al. (2012). These values can be compared with a recent determination of the local dust mass density for galaxies in all environments from Dunne et al. (2011). This shows that Fornax is over dense in dust by about a factor of 120(196) while Virgo is over dense by about a factor of 39(126). 

If the Fornax HI mass function is peaked, similar to Virgo (Davies et al. 2004, Taylor 2010), the HI gas in these galaxies should provide a good estimate of the cluster total. Using the eight HI masses available (Table 3) we derive the HI mass density shown in Table 4. The value for Virgo is taken from Davies et al. (2012).  Recently, Davies et al. (2011) measured a local 'field' HI mass density of $7.9\pm1.2 \times 10^{7}$ M$_{\odot}$ Mpc$^{-3}$ (see also Martin et al. 2010). Thus the Fornax cluster is over dense in HI, when compared to the field, by a factor of 18(29) and Virgo by 6(18). 

As a check of our assumption that our sample galaxies contain the majority of HI in the volume we have looked at the HIPASS data for Fornax as described in Waugh et al. (2002). There are three galaxies in our Herschel field that have HI detections by Waugh et al., but are not in our sample (a visual inspection shows that they are all detected in the far-infrared, but not at the required flux level). The three HI detections are ESO358-G015, ESO358-G051 and ESO358-G060 and using their distances as listed in NED they amount to just 9\% of the total atomic hydrogen in our sample galaxies.

Comparing stellar mass functions is not quite so straight forward because of uncertainties in the faint end slope of the cluster luminosity function i.e the numbers of faint dwarf galaxies (see Sabatini et al., 2003, and references therein). For the stars in our sample galaxies we obtain the densities given in Table 4. The Virgo value is again taken from Davies et al. (2012). Baldry et al. (2008) recently derived the local galaxy stellar mass function and from it the field stellar mass density (Table 4). Thus the Fornax cluster is over dense in stars by 120(195) compared to values for Virgo of 34(129).  Where required in the above derivations we have used H$_{0}$=72 km s$^{-1}$, $\Omega_{m}=0.27$, $\Omega_{\Lambda}=0.73$.

The larger stellar mass density we calculate for Fornax compared to Virgo is again roughly consistent with the factor of 2.5 higher number density of galaxies derived by Ferguson (1989) - Fornax is a higher over density than Virgo. Interestingly, both Virgo and Fornax represent about equal over densities of dust and stars, but the atomic gas is relatively much more depleted (Table 4). This presumably indicates that atomic gas has been removed from these galaxies without the loss of consummate amounts of dust, though Cortese et al. (2010) show, by the clear truncation of Virgo cluster dust discs, that some dust must be lost from the outer regions of galaxies in the cluster environment.

We can make a simple calculation to see the effect of ram pressure on both the gas and the dust. The force per unit area a particle in the disc of a galaxy experiences is the difference between the ram pressure and the restoring force per unit area due to the discs surface mass density (Gunn and Gott, 1972). In this case the acceleration a particle undergoes depends on its cross-sectional area ($a$) and its mass ($m_{p}$) and for constant acceleration the timescale for removal is just proportional to $\sqrt{m_{p}/a}$. Using typical values for the masses ($m_{g}\approx10^{-27}$, $m_{d}\approx10^{-14}$) and sizes ($a_{g}\approx10^{-20}$, $a_{d}\approx10^{-12}$) of gas and dust particles leads a timescale for dust removal about a factor of 100-1000 longer than gas removal (this ignores viscous drag between the gas and dust, see Davies et al. 1998). Evidence from Virgo shows
that atomic and molecular gas are also affected very differently by the cluster environment. There is no evidence for a H$_{2}$ deficiency in Virgo but there clearly is an atomic gas deficiency (Gavazzi et al. 2008, Young et al. 2011).
So, a complementary explanation for the non-removal of dust is that it is primarily associated with dense H$_{2}$ reservoirs
that are very resistant to the ram pressure stripping process.

Using the Davies et al. (2010) and Dunne et al. (2011) values for the global HI and dust mass densities gives a local 'field' atomic gas to dust ratio of 363\footnote{The local value of the gas-to-dust ratio for the Milky Way is $\sim143$ (Draine et al., 2007), but our calculation neglects the molecular gas included in this value.},
while the same ratio in the Fornax and Virgo clusters is 52(53) and 58(50) respectively. From the above it seems that these low values are entirely due to the loss of atomic gas. In each case this amounts to about six times as much HI lost from these galaxies as is presently contained within them ($5 \times 10^{11}$ M$_{\odot}$ and $1 \times 10^{11}$ M$_{\odot}$ respectively for Virgo and Fornax). It seems that irrespective of the different environments prevalent in these two clusters the atomic gas loss mechanisms are equally effective.
 HI complexes within the Virgo cluster, but external to previously identified galaxies, have been found by Davies et al. (2004), Oosterloo and van Gorkom (2005) and Kent et al. (2007), but they only represent a small fraction of the cluster atomic gas. 

\subsection{The spectral energy distribution of NGC1365}
The brightest Virgo cluster galaxy at these wavelengths is NGC4254 ($F_{100}=114.2$ Jy, $D=17$ Mpc, Davies et al. 2012), but this is totally outshone by the Fornax cluster galaxy NGC1365 ($F_{100}=213.5$ Jy, $D=17.9$ Mpc).  
Being so bright NGC1365 is also detected by Planck in 4 bands at 350, 550, 850 and 1382$\mu$m. In Fig 4 we show the PACS, SPIRE and Planck data along with the best fitting modified blackbody taken from Fig 3. The Planck data are those derived by fitting a 2D gaussian
to the data (GAUFLUX, which should be more appropriate for
extended sources) and color corrected using the Herschel
SED and the tables in the explanatory notes of the Planck early release compact source catalogue. The Planck data fit remarkably well the modified blackbody all the way out to 1382$\mu$m. There is no significant sign of any excess emission above that of the modified blackbody, with the point at 1382$\mu$m being the most discrepant. 

The band centred on 1382$\mu$m contains the CO(2-1) line and this might explain the slight excess.
According to Sandqvist et al. (1995)
the total CO(1-0) luminosity of NGC1365 is 5200 Jy km/s. 
For the Planck bandwidth (33\% of the central frequency) and using
the measured CO(2-1)/CO(1-0)=0.8 ratio we find that the CO(2-1) line contributes
0.045 Jy (7\% of the total) to the 1382$\mu$m flux density. If this is subtracted from the color corrected data, than the 1382$\mu$m point sits nicely on top of the
modified blackbody curve. 

Over these longer, Planck wavelengths, we might also expect an additional contribution to the observed flux density from free-free emission (synchrotron emission only starts to become important at wavelengths greater than $\sim$1cm, Condon, 1992). Combining equations 23 and 26 from Condon (1992) we get: \\
\begin{center}
$\left( \frac{F_{\lambda}}{Jy}\right) = 1.2 \times 10^{-10} \left( \frac{\lambda}{\mu m}\right) ^{0.1} \left( \frac{d_{Mpc}}{Mpc}\right)^{-2}\left( \frac{L_{FIR}}{L_{\odot}}\right)$ 
\end{center}
where $F_{\lambda}$ is the free-free flux density in Jy at wavelength $\lambda$ in $\mu$m, $d_{Mpc}$ is the distance in Mpc and $L_{FIR}$ the far-infrared luminosity in solar units. For NGC1365 at a distance of $d_{Mpc}=17.9$ Mpc and $L_{FIR}=L_{100-500}=6.5\times10^{10}$ $L_{\odot}$ the predicted flux densities at 850 and 1382$\mu$m are 0.048 and 0.050 Jy respectively. These values are again much smaller than the extrapolation of the modified blackbody curve and so free-free emission is predicted to be only a small fraction of the emission at these wavelengths.

NGC1365 is a powerful Seyfert (type 1.8) galaxy (Veron-Cetty and Veron 2006), but no sign of the AGN is seen in its global far-infrared/sub-mm spectrum. Previous observations of some more passive galaxies has revealed an excess at wavelengths beyond 350$\mu$m, particularly the Milky Way (Leitch et al. 1997) and star forming dwarf galaxies (Grossi et al. 2010). This emission has been associated with an additional 'spinning dust' component (Draine and Lazarian, 1998a, 1998b), but it is not observed here.

\begin{figure}
\centering
\includegraphics[scale=0.43]{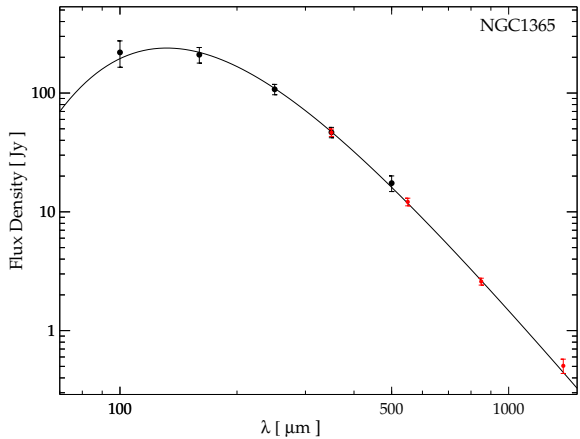}
\caption{The far-infrared spectral energy distribution of NGC1365. Black points are the PACS and SPIRE data and the black line the best modified blackbody curve fit to them. The red points are the Planck data.} 
\end{figure}

\subsection{Emission from NGC1399}
With a B band absolute magnitude of -20.7 NGC1399 is just about a typical ($L^{*}$) galaxy in terms of its optical luminosity. But, being an elliptical galaxy (E1) lying at the very heart of the cluster we might expect that NGC1399's inter-stellar medium is somewhat depleted compared to a galaxy like the Milky Way. No atomic hydrogen has been detected in NGC1399, though there have been previous detections in the far-infrared possibly associated with dust emission. There is an IRAS detection at 0.3 Jy (NED) which would be a marginal detection in our data dependent on the size of the source - but we detect nothing. There is a {\it Spitzer} (MIPS) detection at 160$\mu$m of 0.026 Jy (Temi et al. 2009), but this is below our detection limit. For the similar, though slightly brighter in the optical M87, our observations gave detections in all the PACS/SPIRE bands (Davies et al. 2012), though this was almost exclusively synchrotron emission and not thermal emission from dust (Baes et al. 2010). For synchrotron emission from NGC1399 we expect the highest flux density at the longest wavelengths, consistent with our detection at 500$\mu$m and our non-detection at 100$\mu$m. For the same spectral slope as M87\footnote{There are actually 10 radio flux densities for NGC1399 listed in NED, but there is large scatter in the values with as much as a factor of 7 difference at the same frequency, so we have not used this data to derive a spectral slope.} we would expect a 100$\mu$m flux density of about 0.1 Jy, below our detection threshold and rather at odds with the IRAS 100$\mu$m detection (which is too high) and similarly at odds with the MIPS measurement (which would be too low). 

There are radio synthesis observations (44\arcsec resolution) of NGC1399 at 35cm that clearly show two opposing radio jets emanating from the nucleus (Jones and McAdam, 1992). In Fig. 5 we show the 500$\mu$m image of NGC1399 with the position and extent of the 35cm radio jet indicated by the red arrow. The 500$\mu$m emission appears more extended than the jet and also the jet seems to be off-set from it, but roughly at the same position angle. The optical centre of the galaxy (NED) corresponds with the centre of the jet. For M87 the far-infrared emission clearly traces the radio jet and so it is a bit of a puzzle as to why NGC1399 is different. If the emission is actually due to dust, and this is some form of dust lane, then it is strange that it is not detected more strongly in the other SPIRE bands. It also extends over some 3 arc min (16 kpc) - this is large for a dust feature in an elliptical galaxy - NGC1399 does not appear in the catalogue of dusty ellipticals (Ebneter and Balick, 1985). There is also the possibility that this is some chance alignment of background sources. The region shown in Fig. 5 is some 20 arc min across, specifically made this large to illustrate the brightness and number density of sources across the field. There are background sources that could conspire to produce the emission we are associating with NGC1399, and so at the moment the origin of this 500$\mu$m emission is not clear.

\begin{figure}
\centering
\includegraphics[scale=0.27]{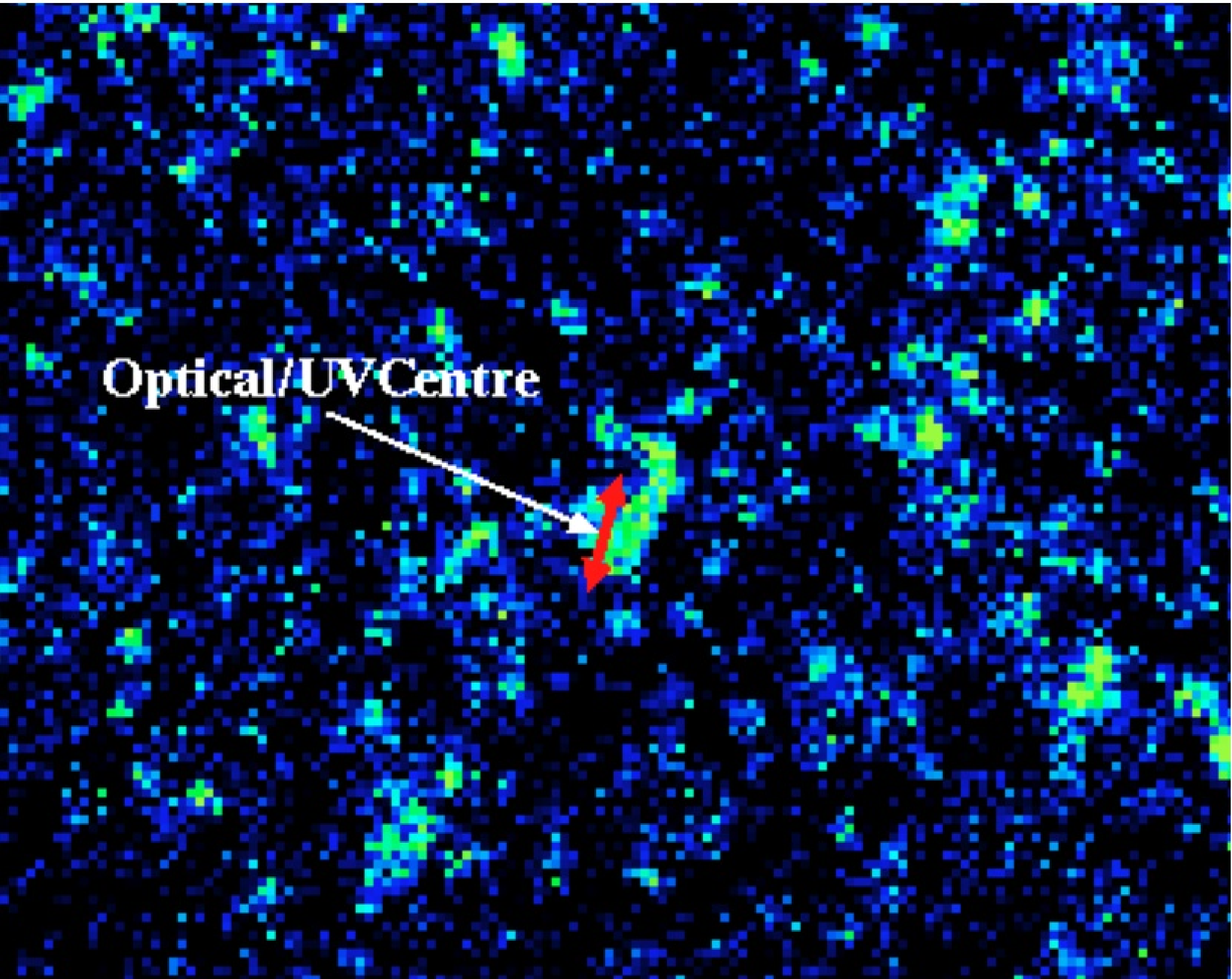}
\caption{ A part of the 500$\mu$m image (radius $\sim22.5$ arc min) centred on the position of NGC1399. Extended emission is seen at approximately the position of NGC1399. The red arrow shows approximately the position, extent and orientation of the radio jet. The centre of the galaxy when measured in the optical and UV is indicated by the white arrow.} 
\end{figure}

\section{Summary}
In this paper we have described the Herschel Fornax Cluster Survey (HeFoCS) and far-infrared observations of 11 bright cluster galaxies. We carry out aperture photometry and show that our measurements are consistent with previous data where available. For 10 of the 11 galaxies we fit spectral energy distribution and derive dust masses and temperatures. We combine dust masses with star and gas masses from the literature to assess the global properties of the galaxies in our sample. These global properties are similar to those detected by us in the Virgo cluster (HeViCS, Davies et al. 2012). We have then compared Fornax and Virgo with each other and the field. Generally Fornax is more dense in stars, gas and dust by a factor of 3-4 relative to Virgo. When compared to the lower density local field of galaxies, Fornax is overdense by a factor of about 120 in stars and dust, but only a factor of 18 in gas. Values for Virgo are about 34 and 39 for stars and dust, reducing to 6 for gas. Both the Fornax and Virgo clusters are relatively depleted in gas to the same extent - gas loss does not seem to be dependent on these two different galactic environments. We specifically discuss two of the sample galaxies in more detail. NGC1365 because it is the brightest far-infrared galaxy in Virgo and Fornax combined. NGC1399 because it is only detected at the longest wavelength of 500$\mu$m. 

\vspace{0.5cm}
\noindent
{\bf ACKNOWLEDGEMENTS} \\

The Herschel spacecraft was designed, built, tested, and launched under a contract to ESA managed by the Herschel/Planck Project team by an industrial consortium under the overall responsibility of the prime contractor Thales Alenia Space (Cannes), and including Astrium (Friedrichshafen) responsible for the payload module and for system testing at spacecraft level, Thales Alenia Space (Turin) responsible for the service module, and Astrium (Toulouse) responsible for the telescope, with in excess of a hundred sub-contractors.

PACS has been developed by a consortium of institutes led by MPE (Germany) and including UVIE (Austria); KU Leuven, CSL, IMEC (Belgium); CEA, LAM (France); MPIA (Germany); INAF-IFSI/OAA/OAP/OAT, LENS, SISSA (Italy); IAC (Spain). This development has been supported by the funding agencies BMVIT (Austria), ESA-PRODEX (Belgium), CEA/CNES (France), DLR (Germany), ASI/INAF (Italy), and CICYT/MCYT (Spain).

SPIRE has been developed by a consortium of institutes led by Cardiff University (UK) and including Univ. Lethbridge (Canada); NAOC (China); CEA, LAM (France); IFSI, Univ. Padua (Italy); IAC (Spain); Stockholm Observatory (Sweden); Imperial College London, RAL, UCL-MSSL, UKATC, Univ. Sussex (UK); and Caltech, JPL, NHSC, Univ. Colorado (USA). This development has been supported by national funding agencies: CSA (Canada); NAOC (China); CEA, CNES, CNRS (France); ASI (Italy); MCINN (Spain); SNSB (Sweden); STFC (UK); and NASA (USA).

This research has made use of the NASA/IPAC Extragalactic Database (NED) which is operated by the Jet Propulsion Laboratory, California Institute of Technology, under contract with the National Aeronautics and Space Administration. 

This publication makes use of data products from the Two Micron All Sky Survey, which is a joint project of the University of Massachusetts and the Infrared Processing and Analysis Center/California Institute of Technology, funded by the National Aeronautics and Space Administration and the National Science Foundation.

The development of Planck has been supported by: ESA; CNES and CNRS/INSU-IN2P3-INP (France); ASI, CNR, and INAF (Italy); NASA and DoE (USA); STFC and UKSA (UK); CSIC, MICINN and JA (Spain); Tekes, AoF and CSC (Finland); DLR and MPG (Germany); CSA (Canada); DTU Space (Denmark); SER/SSO (Switzerland); RCN (Norway); SFI (Ireland); FCT/MCTES (Portugal); and The development of Planck has been supported by: ESA; CNES and CNRS/INSU-IN2P3-INP (France); ASI, CNR, and INAF (Italy); NASA and DoE (USA); STFC and UKSA (UK); CSIC, MICINN and JA (Spain); Tekes, AoF and CSC (Finland); DLR and MPG (Germany); CSA (Canada); DTU Space (Denmark); SER/SSO (Switzerland); RCN (Norway); SFI (Ireland); FCT/MCTES (Portugal); and PRACE (EU).

S. B., L. H. and S. D. A.  acknowledge financial support by ASI through the ASI-INAF grant "HeViCS: the Herschel Virgo Cluster Survey" I/009/10/0.

This work received support from the ALMA-CONICYT Fund for the
Development of Chilean Astronomy (Project 31090013) and from the
Center of Excellence in Astrophysics and Associated Technologies (PBF
06).

P.S.
is a NWO/Veni fellow.

\vspace{0.5cm}
\noindent
\large
{\bf References} \\
\small
Aniano et al., 2012, ApJ, in press (arXiv:1207.4186)  \\
Auld et al., 2012, MNRAS, in press (arXiv:1209.4651) \\
Baes M., Clemens M., Xilouris E., et al., 2010, A\&A, 518, 53 \\
Baldry K., Glazebrook K. and Driver S., 2008, MNRAS, 388, 945  \\
Bell E., McIntosh Daniel H., Katz N. and Weinberg M., 2003, ApJS, 149, 289 \\
Bertin E. and Arnouts S., 1996, A\&ASS, 117, 393 \\
Binggeli B., et al., 1985, AJ, 90, 1681  \\
Boselli A. et al., 2012, A\&A, 540, 54  \\
Boselli A. et al., 2010, A\&A, 518, 61 \\
Boselli A. and Gavazzi G., 2006, PASP, 118, 517  \\
Ciesla L., et al., 2012, A\&A, 543, 161 \\
Clemens M., et al., 2010, A\&A, 518, 50  \\
Condon J., 1992,ARAA, 30, 575 \\
Corbelli E. et al., 2012, A\&A, 542A, 32  \\
Cortese L., et al., 2010, A\&A, 518, 49 \\
Dale D. et al., 2012, ApJ, 745, 95  \\
Davies J., et al., 2012, MNRAS, 419, 3505 \\
Davies J., et al., 2011, MNRAS, 415, 1883 \\
Davies J., et al., 2010, A\&A, 518, 48 \\
Davies J., et al., 2004, MNRAS, 349, 922 \\
Davies J., Alton P., Bianchi S. and Trewhella M., 1998, MNRAS, 300, 1006 \\
de Looze et al., 2010, A\&A, 518, 54  \\
Devereux N. and Young J., 1990, ApJ, 359, 42  \\
Dowell C., et al., 2010, Proc. SPIE 7731, 773136 \\
Draine B. and Lazarian A., 1998a, ApJ, 494, L19 \\
Draine B. and Lazarian A., 1998b, ApJ, 508, 157 \\
Draine B., et al., 2007, ApJ, 663, 866 \\
Drinkwater et al., 1999, ApJ, 511, 97 \\
Drinkwater M., Gregg M. and Colless M., 2001, ApJ, 548, 139 \\
Doyon R. and Joseph R., MNRAS, 1989, 239, 347  \\
Dunne L. et al., 2011, MNRAS, 417, 1510 \\
Ebneter K. and Balick B., 1985, AJ, 90, 183 \\
Ferguson H., 1989, AJ, 98, 36 \\
Ferguson H., 1989a, Ap\&SS, 157, 227  \\
Freedman W., et al., 2001, ApJ, 553, 47  \\
Gavazzi G., et al., 2008, A\&A, 482, 43  \\
Griffin M. et al., 2010, A\&A, 518, 3 \\
Griffin M. et al., 2009, EAS, 34, 33G \\
Grossi M. et al., 2010, A\&A, 518, 52  \\
Gunn J. and Gott R., 1972, ApJ, 176, 1 \\
Horellou C., Casoli F. and Dupraz, C., 1995, A\&A, 303, 361  \\
Jones P. and McAdam W., 1992, ApJS, 80, 137 \\
Jordan C. et al., 2007, ApJSS, 169, 213 \\
Kent B. et al., 2007, ApJ, 665, 15  \\
Leitch E. et al., 1997, ApJ, 486, L23 \\
Martin A., Papastergis E., Giovanelli R., Haynes M., Springob, C. and Stierwalt S., 2010, ApJ, 723, 1359 \\
Magrini L. et al., 2011, A\&A, 535, 13  \\
Mueller T., Nielbock M., Balog Z., Klaas U. and Vilenius E.,
PACS Photometer - Point-Source Flux Calibration
PACS Herschel document PICC-ME-TN-037, v. 1.0 (April 12, 2011) \\
Popescu C. and Tuffs R., 2002, MNRAS, 567, 221 \\
Oosterloo T. and van Gorkom, 2005, A\&A, 437, 19O  \\
Ott S., 2010, In 'Astronomical Data Analysis Systems', ASPC, Ed. Y. Mizumoto, K. Morita and M Ohishi, 434, 139O \\
Paolillo M. et al., 2002, ApJ, 565, 883 \\
Pappalardo et al., 2012, A\&A, 545, 75 \\
Pilbratt G. et al., 2010, A\&A, 518, 1 \\
Poglitsch A. et al., 2010, A\&A, 518, 2 \\
Roussel H., 2012, arXiv:1205.2576 \\
Sabatini S. et al.,  2003, MNRAS, 341, 981  \\
Sandqvist A., Joersaeter S. and Lindblad P., 1995, A\&A, 295, 585 \\
Saunders W. et al., 1990, MNRAS, 242, 318  \\
Schindler S., Binggeli B. and Bohringer, 1999, A\&A, 343, 420  \\
Scharf C. et al., 2005, ApJ, 633, 154 \\
Smith M., et al., 2010, A\&A, 518, 51 \\
Soifer B. et al., 1987, ApJ, 320, 238 \\
Swinyard B. et al., 2010, A\&A, 518, 4 \\
Taylor R., 2010, PhD thesis, Cardiff University, UK. \\
Temi P., Brighenti F. and Matthews W., 2009, ApJ, 707, 890 \\
Thomas P., Drinkwater M. and Evstigneeva E., 2008, MNRAS, 389, 102 \\
Tuffs R. et al., 2002, ApJS, 139, 37  \\
Veron-Cetty M. and Veron P., 2006, A\&A, 455, 773  \\
Waugh M.et al., 2002, MNRAS, 337, 641  \\
Young L. et al., 2011, MNRAS, 414, 940 \\

\end{document}